
\baselineskip=13pt
\magnification=\magstep1

\parskip=\smallskipamount 

\def\quote#1{\medskip
{\baselineskip=12pt{\narrower\narrower\parindent = 0pt #1
\par}\medskip}} 

\def\wquote#1{
{\narrower\parindent = 0pt #1
\par}\medskip}

\def\abst#1{\medskip{\baselineskip=14pt
{\narrower\narrower\parindent = 0pt #1 \par}}} 

\newcount\ftnumber
\def\ft#1{\global\advance\ftnumber by 1
          {\baselineskip 12pt    
           \footnote{$^{\the\ftnumber}$}{#1}}}

\newcount\fnnumber
\def\fn{\global\advance\fnnumber by 1{$^{\the\fnnumber}\,\,$}}

\def\ni{\noindent}

\def\fr#1/#2{{\textstyle{#1\over#2}}}

\def\>{\rangle}
\def\<{\langle}
\def\k#1{|#1\>}
\def\b#1{\<#1|}
\def \ip#1#2{\< #1 | #2 \>}
\def\x{\otimes}
\def\+{+}
\def\s{{\cal S}}
\def\now{{\it now\/}}

\def\tr{{\rm Tr}}

\newcount\eqnumber

\def\eq(#1){
    \ifx\DRAFT\undefined\def\DRAFT{0}\fi	%if undef'd, make it 0
    \global\advance\eqnumber by 1%
    \expandafter\xdef\csname !#1\endcsname{\the\eqnumber}%
    \ifnum\number\DRAFT>0%
	\setbox0=\hbox{#1}%
	\wd0=0pt%
	\eqno({\offinterlineskip
	  \vtop{\hbox{\the\eqnumber}\vskip1.5pt\box0}})%
    \else%
	\eqno(\the\eqnumber)%
    \fi%
}
\def\(#1){(\csname !#1\endcsname)}

%\rightline{ \number\month /\number\day /\number\year}
%; (\number\time)} 

\centerline{{\bf What Is Quantum Mechanics Trying to Tell Us?}\fn}
\bigskip
\centerline{N. David Mermin}
\centerline{Laboratory of Atomic and Solid State Physics}
\centerline{Cornell University, Ithaca, NY 14853-2501}
\bigskip
\abst{{\baselineskip = 10pt I explore whether it is possible to make
sense of the quantum mechanical description of physical reality by
taking the proper subject of physics to be correlation and only
correlation, and by separating the problem of understanding the
nature of quantum mechanics from the hard problem of understanding the
nature of objective probability in individual systems, and the even
harder problem of understanding the nature of conscious awareness. The
resulting perspective on quantum mechanics is supported by some
elementary but insufficiently emphasized theorems.  Whether or not it
is adequate as a new {\it Weltanschauung}, this point of view toward
quantum mechanics provides a different perspective from which to teach
the subject or explain its peculiar character to people in other
fields.}} 

\vfil\eject

\centerline{{\bf What Is Quantum Mechanics Trying to Tell Us?}$^1$}
\bigskip

\quote{{\sl [W]e cannot think of {\rm any} object apart from the
possibility of its connection with other things.} Wittgenstein, {\it
Tractatus, 2.0121}}

\quote{{\sl If everything that we call ``being'' and ``non-being'' consists
in the existence and non-existence of connections between elements, it
makes no sense to speak of an element's being (non-being)$\ldots\,.$} 
Wittgenstein, {\it Philosophical Investigations\/}, 50.}

\quote{{\sl It happened to him as it always happens to those who turn to
science $\ldots$ simply to get an answer to an everyday question of
life. Science answered thousands of other very subtle and ingenious
questions $\ldots$ but not the one he was trying to solve.}  Tolstoy,
{\it Resurrection\/}, XXX.}

\quote{{\sl [I]n our description of nature the purpose is not to
disclose the real essence of the phenomena but only to track down, so
far as it is possible, relations between the manifold aspects of our
experience.}  Bohr.\fn} 

\bigskip
\leftline{{\bf I. What quantum mechanics is trying to tell us}}
\medskip

I would like to describe an attitude toward quantum mechanics which,
whether or not it clarifies the interpretational problems that
continue to plague the subject, at least sets them in a rather
different perspective.  This point of view alters somewhat the
language used to address these issues --- a glossary is provided in
Appendix C --- and it may offer a less perplexing basis for teaching
quantum mechanics or explaining it to non-specialists.  It is based on
one fundamental insight, perhaps best introduced by an analogy.

My complete answer to the late 19th century question ``what is
electrodynamics trying to tell us'' would simply be this: 
\quote{{\it Fields in empty space have physical reality; the medium
that supports them does not.\/}} 
\ni Having thus removed the mystery from electrodynamics, let me immediately
do the same for quantum mechanics: 
\quote{{\it Correlations have physical
reality; that which they correlate does not.\/}}  
\ni The first proposition probably sounded as bizarre to most late
19th century physicists as the second sounds to us today; I expect
that the second will sound as boringly obvious to late 21st century
physicists as the first sounds to us today.  

And that's all there is to it.  The rest is commentary.

\bigskip
\leftline{{\bf II. Correlations and only correlations}}
\medskip

Let me expand on my ten-word answer to what quantum mechanics is all
about, which I have called elsewhere\fn the Ithaca interpretation of
quantum mechanics (IIQM).

Note first that the term ``physical reality'' is not necessarily
synonymous with unqualified ``reality''.  The distinction is of no
interest in understanding what classical electrodynamics is trying to
tell us, but it may be deeply relevant to why quantum mechanics has
not been widely seen to be a theory of correlation without correlata.
I shall set aside for now the tension between {\it reality\/} and {\it
physical reality\/}, but as noted in Section IV below, it will come
back to force itself upon us.\fn

According to the IIQM the only proper subjects for the physics of a
system are its correlations.  The physical reality of a system is
entirely contained in (a) the correlations among its subsystems and
(b) its correlations with other systems, viewed together with itself as
subsystems of a larger system.  I shall refer to these as the {\it
internal\/} and {\it external\/} correlations of the system.  A {\it
completely isolated\/} system is one that has no external
correlations or external dynamical interactions.  

The wave function of a physical system (when it has one) or, more
generally, its quantum state (pure or mixed) is nothing more than a
concise encapsulation of its internal correlations.  Insofar as the
state or the wave function (when the state is pure) has physical
reality, that reality does not extend beyond the reality of the
internal correlations that the state encodes.  In this respect the
IIQM agrees with Bohr and Heisenberg, who viewed the wave function as
nothing more than a computational tool.  It disagrees with
Schr\"odinger's early view of the wave function, or with the views of
currently active deviant subcultures, such as the Bohm-de Broglie
interpretation,\fn and its recent refinements, or
efforts to modify quantum mechanics by making wave function
``collapse'' a dynamical physical process.\fn 

The IIQM does not emerge from a general view of the world out of which
quantum mechanics is extracted; the strategy is rather to take the
formalism of quantum mechanics as given, and to try to infer from the
theory itself what quantum mechanics is trying to tell us about
physical reality.  Thus by systems and subsystems I simply mean the
conventional representation of a complex system by products of
subsystem state spaces. If the system, for example, is a Heisenberg
model of a number of magnetic ions, the subsystems are the spin
degrees of freedom of the individual ions.  If the system is a
hydrogen atom, the subsystems could be the electron and the proton,
further resolved, if this is of interest, into their spin and orbital
degrees of freedom.  In an example that preoccupied the founders of
the theory, the system is an experiment, and the subsystems are the
microscopic object of study and the macroscopic apparatus used to
study it.

The crucial formal property of a resolution into subsystems is that
all observables associated with one subsystem must commute with all
observables associated with any other distinct subsystem.  So if the
subsystems are interacting, then we are dealing with subsystem
correlations at a given time.  A further requirement is that that the
subsystem subspaces whose product makes up the state space for the
entire system can be straightforwardly identified in the standard way
with physically meaningful subsystems of a real (or model) physical
system --- i.e.  that the resolution into subsystems is in some sense
natural, as it is in the above examples.\fn

By correlations among subsystems I have in mind the mean values, at any
given time, of all system observables (hermitian operators) that
consist of products over subsystems of individual subsystem
observables.  Among the observables of a subsystem are the projection
operators onto its linear subspaces, so the set of all correlations
among the subsystems contains the set of all joint probability
distributions over subsystems. Since these distributions are in turn
enough to determine the means of the products of all observables, it
does not matter whether one interprets ``correlations'' to mean joint
distributions, or means of products of observables.  I shall use
whichever interpretation is more appropriate to the case at hand, but
I should emphasize that I use the term ``correlation'' in a sense in
which the absence of correlation (arising when a joint distribution
factors) is regarded as correlation of a degenerate ({\it trivial\/})
form.

It is a remarkable (but not often remarked upon) feature of the
quantum mechanical formalism that all the joint distributions
associated with any of the possible resolutions of a system into
subsystems and any of the possible choices of observable within each
subsystem, are mutually compatible: they all assign identical
probabilities within any sets of subsystems to which they can all be
applied.\fn The physical reality of subsystem correlations need
therefore not be restricted to any particular resolution of a system
into subsystems or to particular choices of observable within each
subsystem, even though different observables for a given subsystem
fail, in general to commute.  It is only when one tries to go beyond
their inter-subsystem {\it correlations\/} to actual {\it correlata\/}
--- particular {\it values\/} for the subsystem observables --- that
non-commuting observables are incapable of sharing simultaneous
physical reality.\fn

The central conceptual difficulty for the IIQM is
the puzzle of what it means to insist that correlations {\it and
only\/} correlations have physical reality.  The ``and only'' part is
an inescapable consequence of many different ``no-hidden-variables''
theorems, as discussed in Section IX below.  These theorems require
that if all correlations have simultaneous physical reality, then all
the correlated quantities themselves cannot.  This problem --- how to
make sense of correlations without correlata --- brings us up against
two major puzzles: 

\wquote{(1) How is probability to be understood as an intrinsic
objective feature of the physical world, rather than merely as a
tactical device for coping with our ignorance?  How is one to make
sense of fundamental, irreducible correlation?}

\wquote{(2) Physics, at least as we understand it today, has nothing to say
about the phenomenon of consciousness.  Conscious reality has more
content than physical reality.}  

I propose to set aside both of these puzzles.  Many of the
difficulties one encounters in interpreting quantum mechanics stem
from our inadequate understanding of objective probability and of
conscious awareness.  It seems worth inquiring whether one can make
sense of quantum mechanics {\it conditional\/} on eventually making
sense of these two even more difficult problems.  I shall therefore
take the notion of {\it correlation\/} as one of the primitive
building blocks from which an understanding of quantum mechanics is to
be constructed.  And I shall take the extraordinary ability of
consciousness to go beyond its own correlations with certain other
subsystems to a direct perception of its own underlying correlata as a
deep puzzle about the nature of consciousness, that ought not,
however, to be a stumbling block in constructing an understanding of
the quantum mechanical description of the non-conscious world.

Before moving to the effort to make sense of quantum mechanics, let me
expand on the two puzzles to be set aside.

\bigskip\leftline{{\bf III. The puzzle of objective
probability.}}\nobreak\medskip\nobreak If correlations constitute the {\it
full\/} content of physical reality, then the fundamental role
probability plays in quantum mechanics has nothing to do with
ignorance.  The correlata --- those properties we would be ignorant of
--- have no physical reality.  There is nothing for us to be ignorant
of.  

A probability that deals only with correlation cannot be based on an
ensemble of copies of a given system, with properties having definite
values in each copy, for the physical absence of correlata applies
separately to each copy.  The only physical description it is possible
to give each individual member of such an ensemble, is in terms of its
own internal correlations.  There is thus no physical or
conceptual role for such an ensemble to play.  All its members are
physically identical, each completely characterized by the identical
set of internal probabilities.  The appropriate context for a theory
of correlations without correlata is one in which probabilistic
notions have meaningful application to individual systems.

It is entirely appropriate for a physics that is both fundamental and
probabilistic to apply directly to individual systems.  The natural
world, after all, consists of individual systems; ensembles are an
artificial contrivance or, at best, a very special kind of composite
individual system.  One motivation behind the desire for an ensemble
interpretation of quantum probabilities is a yearning (not always
acknowledged) for hidden variables (of which values for correlata
constitute the most important example).  The view that probabilistic
theories are about ensembles implicitly assumes that probability is
about ignorance; the hidden variables include whatever it is we are
ignorant of.  But in a non-deterministic world, probability has
nothing to do with incomplete knowledge.  Quantum mechanics is the
first example in human experience where probabilities play an
essential role even when there is nothing to be ignorant
about.  The correlations quantum mechanics describes prevail among
quantities whose individual values are not just unknown: they have no
physical reality.  We lack an adequate understanding of how
probability or correlation is to be understood under such conditions,
but ensemble interpretations fail to capture this central
feature.

Another motivation for an ensemble interpretation of quantum
probability is the intuition that because the {\it predictions\/} of
quantum mechanics are fundamentally probabilistic rather than
deterministic, quantum mechanics only can make sense as a theory of
ensembles.  Whether or not this is the only way to understand
probabilistic predictive power, physics ought to be able to {\it
describe\/} as well as {\it predict\/} the behavior of the natural
world.  The fact that physics cannot make a deterministic prediction
about an individual system does not excuse us from pursuing the goal
of being able to construct a description of an individual system at
the present moment, and not just a fictitious ensemble of such systems.

I shall not explore further the notion of probability and correlation
as objective properties of individual physical systems, though the
validity of much of what I say depends on subsequent efforts to make
this less problematic.  My instincts are that this is the right order
to proceed in: objective probability arises {\it only\/} in quantum
mechanics.  We will understand it better only when we understand
quantum mechanics better.  My strategy is to try to understand quantum
mechanics contingent on an understanding of objective probability, and
only then to see what that understanding teaches us about objective
probability.\fn  

So throughout this essay I shall treat correlation and probability as
primitive concepts, ``incapable of further reduction $\ldots$ a
primary fundamental notion of physics.''\fn The aim is to
see whether all the mysteries of quantum mechanics can be reduced to
this single puzzle.  I believe that they can, provided one steers
clear of another even greater mystery: the nature of ones own personal
consciousness.  

\bigskip\leftline{{\bf IV. The puzzle of consciousness.}}\nobreak Consciousness
enters the picture through the disquieting but indisputable fact that
{\it I\/} know perfectly well that my individual particular {\it
perceptions\/} of certain kinds of subsystems {\it do\/} have a
reality that goes beyond the correlation my perceptions have acquired
with the subsystem through my interaction with it.  It has become
traditional in this context to call such subsystems classical or
macroscopic.  I {\it know\/} that that photomultiplier
\#1 fired and photomultiplier \#2 did not.  I directly perceive the
particularity of my conscious representation of the photomultipliers
from which I infer the particularity of the photomultiplier
excitations themselves.  

To the extent that ``I'' am describable by physics, which deals only
with the correlations between me and the photomultipliers, physics
can only (correctly) assert that photomultiplier
\#$n$ firing is perfectly correlated with my knowing that
photomultiplier \#$n$ fired for either value of $n$.  The question
that physics does not answer is how it can be that {\it I know\/}
that it {\it is\/} \#1 and {\it is not\/} \#2.  This is indeed a
problem.  It is part of the problem of  consciousness.  

The problem of consciousness is an even harder problem than the
problem of interpreting quantum mechanics, and it is important not to
confuse the two.  As with the puzzle of objective probability, here
too it seems sensible to attempt first to understand quantum mechanics
in full awareness of the fact that we do not understand consciousness,
taking the view that consciousness is beyond the scope of physical
science, at least as we understand it today.  This (and only this) is
why I distinguish between {\it reality\/} and {\it physical
reality\/}.  Physical reality is narrower than what is real to the
conscious mind.  Quantum mechanics offers an insufficient basis for a
theory of everything if everything is to include consciousness.

Before relegating the problem of consciousness to the filing
cabinet of harder problems to be examined after satisfactorily
interpreting quantum mechanics --- we shall be forced on various
occasions in the pages that follow to acknowledge the existence of that
cabinet --- let me note some manifestations even in classical physics
of the ability of consciousness to apprehend what physics
cannot.  

The notion of \now\ --- the present moment --- is immediately
evident to consciousness as a special moment of time (or a brief
interval --- of order perhaps a few tenths of a second).  It seems
highly plausible to me that your
\now\ overlaps with my
\now\ or, if you are very far away from me, with a region space-like
separated from my
\now.  On the other hand, I can conceive of it not working this way
--- that your \now\ is two weeks behind or fifteen minutes ahead of my
\now.  In that case when we have a conversation each of us is talking
to a mindless hulk.  I mention this not because I believe in mindless
hulks but because you encounter them in discussions of the ``many
worlds'' interpretation of quantum mechanics.  I do not believe in
many worlds any more than I believe in many \now s, but I find it
significant that the imagery evoked in thinking about a purely
classical puzzle of consciousness is the same as that encountered in
the many worlds attempt to extend quantum mechanics to account for our
conscious perceptions.\fn 

Physics has nothing to do with such notions.  It knows nothing of
\now\ and deals only with correlations between one time and another.
The point on my world-line corresponding to \now, obvious as it is to
{\it me\/}, cannot be identified in any terms known to today's
physics.  This {\it particularity\/} of consciousness --- its ability
to go beyond time differences and position itself absolutely along the
world-line of the being that possesses it --- has a similar flavor to
its ability to go beyond its own correlations with a subsystem, to a
direct awareness of its own particular correlatum and therefore, by
inference, an awareness of a particular subsystem
property.\fn

An even simpler example of an elementary constituent of consciousness
which physics is silent on, is the quality of the sensation
of {\it blueness\/}.  Physics can speak of a certain class of
spectral densities of the radiation field, it can speak of the
stimulation of certain receptors within the eye, it can speak of nerve
impulses from the eye to the visual cortex, but it is absolutely
silent about what is completely obvious to me (and I assume to you)
--- the characteristic and absolutely unmistakable {\it blue\/} quality
of the experience of blueness itself.\fn  

Consciousness enters into the interpretation of quantum mechanics
because it and it alone underlies our conviction that a purely relational
physics --- a physics of correlations without correlata --- has
insufficient descriptive power.  Consciousness cannot easily be banished from
such discussions, because the conviction arises in contexts where the
underlying conscious perception may only be implicit.\fn  One must
therefore remain aware of its ramifications, as a mystery in its own
right, so one can disentangle the characteristic puzzles of
consciousness from efforts to come to terms with the lesser puzzle of
understanding the quantum mechanical description of the non-conscious
world.\fn

\bigskip \leftline{{\bf V. A Theorem about Quantum Correlations\/}}
\nobreak\medskip\nobreak There is a common-sense appeal to the idea of
a physics that is mute on absolute subsystem properties, restricted in
its scope to the correlations among such properties.  Why should
physics be able to produce more than a description of the world in the
world's own terms, by relating some parts of the world to other parts?
More substantially, it is pertinent to note that I am on firm ground
in insisting that the entire content of the physics of a system
consists of a specification of the correlations among its subsystems,
because this happens to be true.  It is the content of an
insufficiently noted but quite elementary theorem, important enough to
deserve a section of its own.

 It is well known that if you are given the
mean value of {\it all\/} the observables of a system, then this
uniquely determines its quantum state (pure or mixed).  Suppose,
however, that the mean values you are supplied with are restricted to
those of observables that are products of subsystem observables over
some specific resolution of the system into subsystems --- i.e. you
are only supplied with the set of all correlations among a particular
set of subsystems that combine to make up the entire system.  How well
is the state of the whole system pinned down when the set of specified
mean values is restricted to such products over subsystems of
subsystem observables, excluding observables that extend globally
over the entire system?  

The surprising (if you've never thought about it) answer is this:
Completely!  {\it Subsystem correlations (for any one resolution of
the system into subsystems) are enough to determine the state of the
entire system uniquely.} This theorem must have been noticed early on,
but the oldest statements of it that I know of are
improbably recent.\fn I shall refer to it as the Theorem on the
Sufficiency of Subsystem Correlations or SSC Theorem. It follows
immediately from three facts:

\wquote{(1) As noted, the means of {\it all\/} observables for the
entire system determine its state.} 
\wquote{(2) The set of all products over subsystems of subsystem
observables contains a basis for the algebra of {\it all\/} such
system-wide observables.}
\wquote{(3) The algorithm that supplies observables with their mean
values is linear on the algebra of observables.} \ni As a result if
you are given the mean values of all such product-over-subsystem
observables, it is a matter of simple arithmetic to compute the mean
values of whatever set of global system observables you need to pin
down the state.  

This is spelled out in detail in Appendix A.  As a simple example, if
a system consists of two spin-$\fr1/2$ subsystems, then the projection
operator on the singlet state --- the state of zero {\it total\/} spin
--- is a global system observable.  It has the well known form
$$P_{\rm singlet} = \fr1/4 \bigl(1- 
\sigma^{1}_x\x\sigma^{2}_x -
\sigma^{1}_y\x\sigma^{2}_y -
\sigma^{1}_z\x\sigma^{2}_z \bigr),\eq(sigma)$$ and therefore its mean
value is entirely determined by the mean values of the products of the
$x$-, $y$-, and $z$-components of the individual spins.  Since the
singlet state is that unique state in which $P_{\rm singlet}$ has the
mean value 1, the system will be in the singlet state provided these
three quantities all have the value $-1$ that expresses perfect
anti-correlation.  

That like components of the individual spins are perfectly
anti-correlated in the singlet state is a famously familiar fact; that
perfect anti-correlations of three orthogonal components is enough to
ensure that the global state {\it is\/} the singlet state --- a
particularly simple playing out of the possibility guaranteed by
the SSC Theorem --- is not as familiar. 
 
Though the proof of the SSC Theorem is elementary, its conceptual
implications are profound.  If the quantum theoretical description of
the physical reality of a system is complete, then so is the
description of the system entirely in terms of all the correlations
that prevail among any specified set of its subsystems, because the
information contained in either of those two descriptions is the
same.  Anything you can say in terms of quantum states --- and some
strange things can be stated in that language --- can be translated
into a statement about subsystem correlations --- i.e.  about joint
probability distributions.  At a minimum, whether or not the IIQM can
be made into a coherent whole, this simple fact ought to be stressed
in all introductory expositions of the quantum theory: 
\quote{{\it The quantum state of a complex system is nothing more than
a concise encapsulation of the correlations among its
subsystems.\/}}

The quantum state is a remarkably powerful encoding of those
correlations.  It enables us to calculate them for any resolution of
the system into subsystems and for any set whatever of subsystem
observables.  The fact that all the different sets of subsystem
correlations can be encoded in a single quantum state provides an
explicit demonstration of the mutual consistency of the correlations
associated with all of the different ways of dividing a system into
subsystems.  While I am not convinced that this shift in point of view
from quantum state to subsystem correlations eliminates all conceptual
problems from the foundations of quantum mechanics, it does alter how
you look at many of those problems and, I believe, offers a better way
to tell people encountering the subject for the first time what it is
all about.  In Sections VI-X I describe some of the shifts in
perspective that take place when you start taking seriously the notion
that the physics of a system is only about the correlations among its
subsystems.

\bigskip \leftline{{\bf VI.  Elimination of measurement from the
foundations}}\nobreak\medskip\nobreak The notion of ``measurement''
plays a fundamental role in conventional formulations of quantum
mechanics.  Indeed quantum mechanics is often presented as merely an
algorithm that takes you from one measurement (``state preparation''
involves selecting a particular output channel from a measurement
apparatus) to another.  John Bell railed eloquently against this.\fn
Why should the scope of physics be restricted to the artificial
contrivances we are forced to resort to in our efforts to probe the
world?  Why should a fundamental theory have to take its meaning from
a notion of ``measurement'' external to the theory itself?  Should not
the meaning of ``measurement'' emerge from the theory, rather than the
other way around?  Should not physics be able to make statements about
the unmeasured, unprepared world?  \quote{To restrict quantum
mechanics to be exclusively about piddling laboratory operations is to
betray the great enterprise.  A serious formulation will not exclude
the big world outside of the laboratory.}  \ni I argue here that the
very much broader concept of correlation ought to replace measurement
in a serious formulation of what quantum mechanics is all about.\fn

The key to freeing quantum mechanics from the tyranny of measurement
is to note that a measurement consists of the
establishment of a particular kind of correlation between two
particular kinds of subsystems, and to insist that everything
that can be said about the physical reality of the correlations
established in a measurement applies equally well to the correlations
among any subsystems of a quantum system.  If physics is about
correlations among subsystems then it is {\it a fortiori\/} about
measurement.  But to insist that physics is exclusively about
measurement, is unnecessarily to relegate to an inferior ontological
status the more general correlations among arbitrary subsystems.

Expanding on this, let me review in its simplest form the standard
characterization of a measurement.  In a measurement a particular
interaction brings about a particular kind of correlation between two
particular subsystems.  One of the subsystems, the one one wishes to
learn about, is arbitrary, but in many important applications it
describes something on the atomic scale.  Call this subsystem
the {\it specimen\/}.  The other subsystem has enormously many degrees
of freedom, describing a piece of laboratory equipment that includes
some sort of readily readable output (which could be in the
form of a pointer, a digital display, or a print-out.)  It is usually
called the {\it apparatus\/}.  

Initially, at the start of a measurement, the specimen and the
apparatus are uncorrelated: the state of the specimen--apparatus
system is a product state $$\k{I} = \k{s}\x\k{a}.\eq(I)$$ To
measure a specimen observable $S$ with eigenstates $\k{s_i}$ one must
establish an interaction between specimen and apparatus that takes an
initial state $\k{s_i}\x\k{a}$ of the combined system into the
final state $\k{s_i}\x\k{a_i}$ where the $\k{a_i}$ are a set of
orthogonal apparatus states associated with macroscopically
distinguishable scale readings: $$ \k{s_i}\x\k{a} \rightarrow
\k{s_i}\x\k{a_i}.\eq(evolve)$$ Because the transformation \(evolve)
takes orthogonal states into orthogonal states it can indeed be
realized by a unitary transformation --- i.e. as a time development
under a suitable choice of Hamiltonian.  Because unitary
transformations
are linear, if the initial state of the specimen has an expansion
$$\k{s} =
\sum\alpha_i\k{s_i},\eq(specstate)$$ then when the measurement
interaction has completed its action, the state of the system will be
$$\k{F} = \sum
\alpha_i\k{s_i}\x\k{a_i}.\eq(F)$$ A correlation has therefore been
established between specimen and apparatus characterized by the joint
probability distribution $$p(s_i,a_j) = \b F P_{s_i}P_{a_j}\k F =
|\alpha_i|^2\delta_{ij}\eq(joint)$$ (where the $P$'s are the
appropriate projection operators: $P_{s_i} =
\k{s_i}\b{s_i}$,\ \  $P_{a_i} =
\k{a_i}\b{a_i}$).  This joint distribution describes a perfect
correlation between apparatus and specimen states: the probability of
the $j$th apparatus state being associated with the $i$th specimen
state is zero unless $i = j$.  And the overall probability of the
$j$th apparatus state is $\sum_i p(s_i,a_j) = |\alpha_j|^2$, which is
just the probability the Born rule assigns to ``the result of a
measurement of $S$ on a specimen in the state $\k s$ yielding the
value $s_j$.'' 

So a measurement of a specimen observable $S$ is an interaction
between the specimen and the apparatus designed to extend the Born
probabilities from the specimen states $\k{s_i}$ to corresponding
apparatus states $\k{a_i}$.  This is a useful thing to do because
although we humans are incapable of directly perceiving the condition
of a microscopic specimen, we are able to perceive the condition of a
macroscopic apparatus.  Both this ability of ours and its limitation
presumably arise from our having evolved under the selective pressure
of having to deal with macroscopic things like tigers and oranges, but
not (at least at the stage of development when consciousness first
arose) with microscopic things like atoms and molecules.  As noted
above, how we manage this conscious perception is deeply mysterious,
but it should be viewed as a mystery about {\it us\/} and should not be
confused with the problem of understanding quantum mechanics.

The great emphasis even today on the particular kinds of correlation
established in a measurement finds its origins in the early history of
the subject.  In the beginning, when people were groping for an
understanding of microscopic specimens, it was natural to express
everything in terms of the more familiar macroscopic apparatuses with
which they were able to correlate the microscopic specimens, through
measurement interactions.  Measurements produced the only correlations
people felt comfortable with.  Today, three quarters of a century
later, having accumulated a vast body of experience dealing with
microscopic specimens, we have developed enough intuition about them
to contemplate usefully a much broader class of correlations in which
no subsystems are required to be of the macroscopic or ``classical''
kind directly accessible to our perception, and in which the
correlations are neither necessarily of the one-to-one type
established in a measurement nor necessarily restricted to just a pair
of subsystems.  

The emphasis on measurement in conventional formulations of quantum
mechanics, and the accompanying emphasis on a classical domain of
phenomena, ought to be viewed as historic relics.  The classical
domain plays a central role only if one restricts the correlations one
is willing to call physically real, to those between specimens and
apparatuses, where an apparatus is a subsystem large enough that we
can perceive it directly --- i.e. a ``classical'' subsystem.  We ought
by now to have outgrown this point of view.  The bipartite
specimen-apparatus correlations produced by a measurement are not the
only kinds of subsystem correlations worthy of being granted physical
reality.  The quantum theory allows us to contemplate together {\it
all\/} the correlations among arbitrary subsystems, and it is simply a
bad habit not to grant micro-micro-$\,\cdots\,$-micro correlations as much
objective reality as the traditional emphasis on measurements has
granted to micro-macro correlations.  

This reluctance to shift the emphasis from measurement to correlation
lies behind statements one often encounters to the effect that
interactions with its environment are in some not very well specified
way continually {\it measuring\/} a specimen.  This is to characterize
a very general state of affairs by a very special and rather atypical
case.  Interactions with its environment have the precise effect of
correlating a specimen with that environment.  Interactions with a
measurement apparatus correlate a specimen with that apparatus.  In
both cases interaction produces correlation.  In measurements the
interactions are designed so that the correlations that develop have
the particular form \(joint) of special interest to us.  It is only the
reluctance to acknowledge that all correlations are real and objective
--- not just those produced by a measurement --- that leads one to
view the more general specimen-environment correlations in terms of
the more special specimen-apparatus correlations produced in a
measurement.

\bigskip
\leftline{{\bf VII. Elimination of knowledge from the
foundations}} 
\medskip

There has always been talk to the effect that quantum mechanics
describes not the physical world but our knowledge of the physical
world.  This intrusion of human knowledge into physics is
distastefully anthropocentric.  In the IIQM such talk is replaced with
talk about objective correlations between subsystems.  Human knowledge
has intruded for two reasons:

(1) The restriction of attention to the correlations established in
measurements has led to an excessively narrow focus on the correlations
between a specimen and what we {\it know\/} about it (or what our
mechanical surrogate --- the apparatus --- records about
it). 

(2) There is a confusion between the strange and
unprecedented role of probability in the quantum theory as an
objective feature of the physical world, and the older better
understood uses of probability as a practical device for coping with
human ignorance.  Because we understand probability reasonably well in
the latter sense, and have only a glimmering of an understanding of
probability in the more fundamental former sense, it is tempting
incorrectly to interpret probabilistic assertions as statements about
human ignorance or knowledge.  

As an important illustration, consider how people distinguish
between pure and mixed states.  It is often said that a system is in a
pure state if {\it we\/} have maximum {\it knowledge\/} of the system,
while it is in a mixed state if {\it our knowledge\/} of the system is
incomplete.  But from the point of view of the IIQM, {\it we\/} are
simply a particular subsystem, and a highly problematic one at that,
to the extent that our consciousness comes into play.  This
characterization of the difference between pure and mixed states can
be translated into a statement about objective correlation between
subsystems, that makes no reference to us or our knowledge:  

By definition, a system $\s_1$ is in a pure state if all the
correlations among any of its own subsystems can be characterized in
terms of a density matrix that is a projection operator onto a
one-dimensional subspace.  This in turn can be shown (Appendix B) to
be possible if and only if any conceivable larger system $\s =
\s_1 + \s_2$ that contains $\s_1$ as a subsystem has only trivial
correlations (i.e.~only factorizable joint distributions) between its
subsystems $\s_1$ and $\s_2$.  Thus {\it a system is in a pure state
if and only if its internal correlations are incompatible with the
existence of any non-trivial external correlations.\/} 

The absence or presence of non-trivial external correlations is the
objective fact.  The anthropocentrisms simply express the consequences
of this fact for us, should we be told all the internal correlations
of $\s_1$.  It is a another remarkable feature of quantum mechanics
(not shared with classical physics, where external correlations are
always possible) that the totality of all possible internal
correlations is enough to determine whether or not any non-trivial
external correlations are possible.  For Appendix A shows that the
internal correlations of a subsystem are enough to determine its
density matrix; and Appendix B shows that non-trivial external
correlations are possible if and only if that density matrix is not a
one-dimensional projection operator.  To characterize the situation
in which the internal correlations are of the kind that prohibit any
external correlations as a situation in which ``we have maximum
knowledge'' is to let {\it ourselves\/} intrude on a formulation that
has no need of {\it us\/}.  

This intrusion of ``knowledge'' into the distinction between pure and
mixed states can lead to another kind of confusion.  It is a common
error always to view a mixed state as describing a system that is
actually in one of a number of different possible pure states, with
specified probabilities.  While this ``ignorance interpretation'' of
the mixed state can indeed be a useful practical way to describe an
ensemble of completely isolated systems, it entirely misses the deep
and fundamental character of mixed states:  if a system has any external
correlations whatever, then its quantum state cannot be pure.  Pure
states are a rarity, enjoyed only by completely isolated systems.  The
states of externally correlated individual systems are fundamentally
and irreducibly mixed.  This has nothing to do with ``our
ignorance''.  It is a consequence of the existence of objective
external correlation.

\bigskip
\leftline{{\bf VIII. The Measurement Problem}}\medskip\nobreak According
to a conventional view, if a specimen is in a state $$\k{s} =
\sum\alpha_i\k{s_i},\eq(specstate1)$$ then after a measurement of an
observable whose eigenstates are the $\k{s_i}$, the state of the system
discontinuously ``collapses'' to the state $\k{s_i}$ with probability
$|\alpha_i|^2$.  At that point all information contained in the phases
of the amplitudes $\alpha_i$ is irredeemably lost.  The ``measurement
problem'' is the problem of how to reconcile this with the continuous
evolution of the specimen-apparatus system into the final state \(F),
which is clearly still capable of revealing interference effects in
the form of probabilities that do depend on the phases of the
$\alpha_i$.  

According to the IIQM the state of a specimen is just a compact
specification of all its internal subsystem correlations.  To
understand collapse, we should restate it not in terms of the {\it
state\/} of the specimen, but in terms of the specimen's {\it internal
correlations.\/} The physical content of the claim that after the
measurement the system ``is in'' the state $\k{s_i}$ with probability
$|\alpha_i|^2$, is that after the measurement the specimen has the
internal correlations appropriate to the state $\k{s_i}$ with
probability $|\alpha_i|^2$.

When it is put this way any discontinuity vanishes.  For as noted
above, during the course of the measurement interaction the combined
specimen-apparatus system evolves continuously from its uncorrelated
initial state \(I) to the highly correlated final state \(F).  As soon
as any non-trivial correlation develops, the state of the specimen
ceases to be pure, and at the end of the interaction when the whole
system is in the state
\(F), the state of the specimen has continuously evolved into the
mixed state $$\sum|\alpha_i|^2\k{s_i}\b{s_i}.\eq(mix)$$ In this mixed
state the internal correlations of the specimen are identical to what
they would be if it were in the pure state $\k{s_i}$ with probability
$|\alpha_i|^2$ --- i.e. the internal correlations are identical to
those given by the collapse story.

This is another familiar tale.  The IIQM shifts the way it is
sometimes told, by emphasizing that the state of a non-trivially
correlated subsystem is never pure: the state of the specimen evolves
continuously from a pure state through a sequence of mixed states into
the ``post-measurement'' mixed state \(mix) at the moment the
measurement interaction completes its task.  If at that stage one wishes to
regard the state of the specimen as undergoing an abrupt change, it is
at worst a collapse from a mixed state viewed in this fundamental way,
to the same mixed state viewed under the ``ignorance
interpretation''.  Since the internal correlations of the specimen are
exactly the same regardless of which view you take, the collapse, if
one chooses so to regard it, is rather ethereal. 

There is thus no quantum measurement problem for the {\it internal\/}
correlations of the specimen or the apparatus.  After the measurement
interaction is complete their states are {\it exactly\/} --- not just
FAPP\fn --- the conventional post-measurement mixed states, which
reveal no interference effects whatever in any probability
distributions associated entirely with the specimen or entirely with
the apparatus.  These mixed states have evolved from the
pre-measurement pure states in an entirely continuous fashion.\fn 

The measurement problem survives only in the {\it specimen\/}-{\it
apparatus correlations} that hold between specimen and apparatus
observables, both of which differ from those characterized by the
joint distribution \(joint) that the measurement interaction was
designed to produce.  Consider, for example, the specimen observable
$$S_{12} = \k{s_1}\b{s_2} + \k{s_2}\b{s_1} \eq(s12)$$ and the
apparatus observable $$A_{12} = \k{a_1}\b{a_2} +
\k{a_2}\b{a_1}. \eq(s12)$$ In the final state \(F) of the
specimen-apparatus system these have nontrivial correlations
$$\b{F}S_{12}A_{12}\k{F}
 = 2{\rm Re}\alpha_1^*\alpha_2 \eq(s12a12)$$ that depend on the
relative phases of the $\alpha_i$, even though those phases can affect
no {\it internal\/} specimen or apparatus correlations in the state
$\k F$. 

There need be nothing peculiar about the specimen observable
$S_{12}$.  If, for example, the specimen is a two-state system viewed
as a spin-$\fr1/2$ and $\k{s_1}$ and $\k{s_2}$ are the eigenstates of
the $z$-component of spin, then $S_{12}$ is just the $x$-component.
On the other hand the apparatus observable $A_{12}$ is quite bizarre,
since its values $\pm 1$ discriminate between the apparatus being in
either of the two superpositions $\k{a_1} \pm
\k{a_2}$ of states with macroscopically distinguishable scale
readings.  ``Macroscopically'' is, of course, crucial.  Were the
``apparatus'' merely another microscopic spin-$\fr1/2$, then \(s12a12)
would give just the correlation in the two $x$-components.  Under
those conditions there would be little trouble introducing a further
straightforward coupling between specimen and apparatus that undid the
measurement interaction, transforming the perfectly correlated system
state with both subsystems in mixed states back into the entirely
uncorrelated system state with both subsystems back in their initial
pure states.  For the same reasons that classical macroscopic systems
are hard to run backwards, the measurement interaction cannot so
readily be undone when the apparatus is macroscopic.  The apparatus
observables whose correlation with the specimen depend on the critical
phases necessary for the reconstruction of the original state are
correspondingly difficult to realize.  

But {\it in principle\/} it could be done.  This is the measurement
problem.  What makes it so much more vexing than the old classical problem of
irreversibility at the macroscopic level is only what happens when {\it
I\/} get into the story.  When {\it I\/} look at the scale of the
apparatus {\it I know\/} what it reads.  Those absurdly delicate,
hopelessly inaccessible, global system correlations {\it obviously\/}
vanish completely when they connect up with {\it me\/}.  Whether this
is because consciousness is beyond the range of phenomena that quantum
mechanics is capable of dealing with, or because it has infinitely
many degrees of freedom or special super-selection rules of its own, I
would not presume to guess.  But this is a puzzle about consciousness
which should not get mixed up with efforts to understand quantum
mechanics as a theory of subsystem correlations in the non-conscious
world.

It is here that the IIQM comes closest to the many-worlds
extravaganza.\fn Many worlds (or many minds) enter the story only when
the formalism is taken to apply to consciousness itself.  In that case,
even though {\it I know\/} that photomultiplier \#1 fired, this
correlation between me and the photomultipliers is associated with
merely one component of a superposition of states of the
me-photomultipliers system.  There is another component in which {\it
I know\/} that photomultiplier \#2 fired.  If quantum mechanics
applies to my conscious awareness (and if there is no objective
physical process of ``wave-function collapse'') then there is no
evading this, and away we go to Fairyland.  But since there are so
many other aspects of conscious awareness that physics has nothing to
say about, I find it naive to assume that it can sensibly be extended
to account for the characteristic particularity of conscious
experience that takes it beyond the correlations between me and the
objects of my knowledge.  

If we leave conscious beings out of the picture and insist that
physics is only about correlation, then there is no measurement
problem in quantum mechanics.  This is not to say that there is no
problem.  But it is not a problem for the science of quantum
mechanics.  It is an everyday question of life: the puzzle of
conscious awareness.

\bigskip
\leftline{{\bf IX. Absence of Correlata}}
\medskip 

In maintaining that subsystem correlations {\it and only} correlations
have physical reality, I have not been very precise about what ``and
only'' is meant to exclude.  One thing that it does {\it not\/}
exclude is the existence of global probability distributions for an
individual subsystem, since these are special cases of its external
correlations with the observables for all the external subsystems
taken to be identically unity.  Indeed, as remarked upon in Section
II, it is a conceptually remarkable (though analytically trivial)
feature of the quantum mechanical formalism that every one of the many
different joint distributions in which a given subsystem $\s_1$
appears gives exactly the same set of marginal distributions for that
given subsystem.  It does not matter which other subsystems
$\s_2,\ldots,\s_n$ appear in the resolution $\s =
\s_1 + \s_2 +\cdots+ \s_n$ of the full system $\s$ into subsystems,
and it does not matter which observable one choses for each of the
other subsystems.  

This is conceptually remarkable because if one takes the orthodox view
that joint distributions apply only to the results of measurement,
then different joint distributions leading to the same marginal
distribution for $\s_1$ characterize mutually exclusive experimental
arrangements, and it is hard to understand why the marginal
distributions for $\s_1$ should be invariant under such changes.  I
have remarked on this elsewhere.\fn It is not remarkable --- on
the contrary, it is essential for the consistency of the whole point
of view --- if the joint distributions are regarded as characterizing
coexisting aspects (all possible subsystem correlations) of physical
reality.  The price one pays for this broader vision of the nature of
joint distributions is the need to deny physical reality to a
complete collection of correlata underlying all these correlations.

The correlata cannot all have physical reality because in spite of the
existence of all subsystem joint distributions and of unique marginal
distributions for individual subsystems, it is impossible to
construct, in the standard way, a full and mutually consistent set of
{\it conditional\/} distributions from the joint and individual
subsystem distributions.  Let me illustrate this extraordinary feature
of quantum probabilities with what is probably the simplest example of
it, discovered by Lucien Hardy in a rather different
context.\fn

Take a system consisting of two subsystems, each describable by a
two-dimensional state space.  Consider just two non-commuting
observables for each subsystem, named 1 and 2 for one subsystem, and
$1'$ and $2'$ for the other.  Label the two eigenstates of each
observable by the name of the observable and one of the two letters
$R$ (for ``red'') or $G$ (for ``green''), and consider the subsystem
correlations in the system state $$\k{\Psi} \propto
\k{2R,2'R} - \k{1R,1'R}\ip{1R,1'R}{2R,2'R}\eq(hardy)$$ (where $\k{X,Y}$
means $\k X\x\k Y$).  According to the IIQM the $11'$, $22'$, $12'$,
and $21'$ subsystem correlations all have simultaneous physical
reality and indeed, we can compute from
\(hardy) the four joint distributions $p(iX,j'Y)$ where each of $i$
and $j$ can be 1 or 2, and each of $X$ and $Y$ can be $R$ or $G$.

Furthermore, the marginal distributions, characterizing one of the two
systems, $$ p(iX) = p(iX,j'R) + p(iX,j'G)\eq(marg1)$$ and $$ p(j'Y) =
p(iR,j'Y) + p(iG,j'Y)\eq(marg2)$$ are indeed independent of whether
the observable for the other (summed over) system is its \#1 or \#2
observable.  There is therefore no formal obstacle to defining in the
conventional way conditional distributions satisfying
$$p(iX|j'Y)p(j'Y) = p(iX,j'Y)\eq(cond1)$$ and $$p(j'Y|iX)p(iX) =
p(iX,j'Y).\eq(cond2)$$ Yet these conditional distributions are
mutually inconsistent.

The proof of this for the Hardy state \(hardy) is simple.  Inspection
of \(hardy) reveals that $\k{\Psi}$ is constructed to be orthogonal to
the state $\k{1R,1'R}$, and it is also orthogonal to the states
$\k{1G,2'G}$, and $\k{2G,1'G}$, since the $R$ and $G$ eigenstates of
any one subsystem observable are orthogonal.  But $\k{\Psi}$ is not
orthogonal to $\k{2G,2'G}$, since for either subsystem the eigenstates
of the \#2 observable are neither orthogonal to nor identical to those
of the \#1 observable.  Consequently the probabilities $p(1R,1'R),\
p(1G,2'G),$ and $p(2G,1'G)$ are zero, but $p(2G,2'G)$ is not:
$$\eqalign{p(1G,2'G) = p(1R,1'R) &= p(2G,1'G) = 0,\cr
          p(2G,2'G) &\neq 0.}\eq(hardyprobs)$$

The vanishing of $p(1G,2'G)$ requires that $$p(1R|2'G)=1, \eq(c1)$$
the vanishing of $p(1R,1'R)$ requires that $$p(1'G|1R) =1, \eq(c2)$$
and the vanishing of $p(2G,1'G)$ requires that $$p(2R|1'G) =1.
\eq(c3)$$ Combining these, if observable $2'$ has the value $G$, then
\(c1) requires $1$ to have the value $R$, in which case \(c2) requires
$1'$ to have the value $G$, in which case \(c3) requires $2$ to have
the value $R$.  So if $2'$ has the value $G$ then 2 must have the
value $R$:  $$p(2R|2'G) = 1.\eq(c4)$$ But this is inconsistent with
the non-zero value of $p(2G,2'G)$.  The statistics
\(hardyprobs) are incompatible with these straightforwardly
constructed conditional distributions.

The conventional interpretation of quantum mechanics finds the above
line of reasoning unacceptable.  According to the
conventional view, probabilities like $p(iX,j'Y)$ are not measures of
some pre-existing set of objective correlations between all four pairs
of subsystem observables.  These probabilities apply only to the
results of actual measurements.  The probability $p(1R,2'G)$ is the
probability that a joint {\it measurement\/} of observables $1$ and
$2'$ yields the values $R$ and $G$.  The three conditional
distributions \(c1)-\(c3) do not characterize coexisting states of
being, but the results of mutually exclusive experiments.  Since at
most one of the experiments can actually be performed, at most one of
the distributions is meaningful, and it makes no sense to combine them
as I have done.

But the IIQM takes a broader view of joint distributions.  All
correlations among all possible subsystem observables have
simultaneous physical reality.  In particular all four pair
distributions have physical reality, whether or not one chooses to
extend the correlations between a particular one of these pairs to a
pair of apparatuses by means of an appropriately chosen measurement
interaction.  What the preceding argument demonstrates is that if all
the subsystem joint distributions do share a common physical reality,
then the {\it conditional\/} distributions constructed from them
cannot, even though all the joint distributions yield unique mutually
consistent marginal distributions for the subsystems.  But if it makes no
physical sense to talk about the probability of 1 being $R$, {\it
given\/} that $2'$ is G, this can only be because absolute subsystem
properties are not ``given''.  If physical reality consists of all
the correlations among subsystems then physical reality cannot extend
to the values for the full set of correlata underlying those
correlations.

The way we conventionally speak of probability makes it hard to
express this state of affairs.  One tends, for example, to speak of
$p(1R,2'G)$ as the probability that 1 is $R$ and $2'$ is $G$.  But if
it makes sense to speak of 1 {\it being\/} $R$ and $2'$ {\it being\/}
$G$, why should it not make equal sense to speak of the probability of
1 being $R$, {\it given\/} that $2'$ is $G$?  The answer has to be
that $p(1R,2'G)$ cannot be viewed as the probability that 1 $is$ $R$
and $2'$ $is$ $G$.  This would make sense were probability a device
for coping with ignorance, but the objective probabilities of quantum
mechanics exist even though there is nothing to be ignorant of.  They
express correlations in the absence of correlata.  To avoid such
linguistic traps it would be better to speaking not of
``probabilities'' but of ``propensities'' or ``dispositions'', or to
eschew all talk of probability in favor of talk about correlation.

I am not suggesting that banishing ``probability'' from our vocabulary
will remove all puzzles from quantum mechanics; only that it can help
avoid misuses of that term.  As noted in Section III, the problem of
what objective probability or objective correlation or propensity
might mean --- of what it means to have correlation when values cannot
be assigned to the correlata --- is one I propose to set aside to
explore whether one can make better sense of quantum mechanics,
contingent on acquiring a better understanding of this admittedly
peculiar notion.  What Hardy's state \(hardy) tells us is that if all
correlations between subsystems {\it do\/} have joint physical
reality, then distributions conditional on particular subsystem
properties {\it cannot\/} in general exist, and therefore such
correlations {\it must\/} be without correlata.\fn

\bigskip\leftline{{\bf X. Nonlocality?}}\nobreak\medskip\nobreak
Hardy did not come up with the state \(hardy) to demonstrate that the
joint existence of pair distributions is incompatible with the joint
existence of conditional distributions.  He produced it as a succinct
and powerful contribution to the tradition of ``nonlocality''
arguments stemming from Bell's theorem.\fn

Under the IIQM, such arguments do not work as demonstrations of
nonlocality.  If two subsystems are spatially separated then the {\it
local\/} properties of each are limited to their {\it internal\/}
correlations.  These are completely determined by the density matrix
of each.  The density matrix of either subsystem is unaffected by any
dynamical process acting only on the other subsystem, even when the
dynamical process consists of letting the other subsystem undergo a
measurement interaction with a third subsystem that functions as an
apparatus.   The choice and performance of a measurement on one
subsystem cannot alter the local properties of the other, far away
subsystem.  Otherwise one could use ``quantum nonlocality'' to
send instantaneous signals.  The impossibility of doing this should be
called {\it physical locality\/}.

Quantum mechanics obeys physical locality.  ``Quantum nonlocality'' (a
violation, so to speak, of {\it metaphysical locality\/}) arises when
one tries to reconcile the {\it actual\/} results of specific
experiments to the {\it hypothetical\/} results of other experiments
that might have been performed but were not.  In talking of ``actual
results'' one is going beyond the subsystem correlations with which
physics can deal, to our mysterious ability to perceive --- i.e.
become consciously aware of --- a particular one of the correlated
possibilities, when we ourselves are among the subsystems.  While it
is surely unreasonable to insist that we have no right to try to make
sense of our own direct perceptions, this kind of reasoning goes
beyond what can be expressed in proper {\it physical\/} terms.   Pearle.

Nevertheless, the following line of thought has a powerful appeal.
Consider a series of experiments in which two particles interact in
such a way as to leave them in the Hardy state, and then fly apart to
separate measurement apparatuses in a manner that preserves the Hardy
state correlations of all the \#1 and \#2 observables.  This is possible
if those observables are, for example polarizations along different
non-orthogonal directions.  

Consider a series of measurements in which the choice of which
observable to measure is decided by tossing a coin at the site of the
measurement.  Consider a run in which the coin tosses result
in observables 2 and 2$'$ being measured, and in which the result of
each measurement is perceived to be $G$.  (The non-vanishing of
$p(2G,2'G)$ guarantees that such runs are possible.)  Suppose the
measurement interactions take place in space-like separated space-time
regions, so there is a frame of reference (Alice's --- let her be
in the vicinity of the unprimed measurement as it takes place) in
which the perception of $G$ at the unprimed system occurs before the
toss of the coin at the primed system, and another frame (Bob's ---
let him be in the space-time neighborhood of the primed measurement) in
which $G$ is perceived at the primed system before the toss at the
unprimed system. 

Once Alice perceives $G$ for the 2-measurement, she is surely entitled
to conclude that if the (yet to be performed in her frame) coin toss
results in a $1'$-measurement on the primed system, the result will be
perceived to be $R$, since $p(2G,1'G) = 0.$ By the same token once Bob
perceives $G$ for the $2'$ measurement he can correctly conclude that
if the (still unperformed in his frame) toss at the unprimed system
results in a $1$ measurement the perceived result must be $R$.  

How can these two valid conclusions be reconciled with the fact that
$1R$ and $1'R$ are never jointly perceived? There are two options.
The first is to abandon the implicit assumption that the perceived
result of a later measurement is unaffected by the choice and/or
outcome of an earlier one.  This is a route taken by those who
embrace quantum nonlocality.  It has the disconcerting feature that
which measurement process affects which depends on whether you are
using Alice's frame of reference or Bob's, but since the influence is
of one of two space-like separated events on another, this is
unavoidable.  The most determined efforts to extract nonlocality from
this kind of reasoning are those of Henry Stapp,\fn

The second option (which I prefer) is to deny that the combined
predictions of Alice and Bob have any relevance to {\it what would have
been perceived\/} if both measurements had actually been of type 1.
Indeed, it is hard to give ``what would have been perceived'' any
meaning in this case, since both predictions are based on actual
perceptions of type-2 measurements.  Alice, for example, having
perceived $G$ in her type-2 measurement is perfectly correct in
concluding that if the toss of Bob's coin results in a $1'$
measurement then Bob will necessarily perceive $R$.  Similarly for
Bob.  But to extract from this a contradiction with the impossibility
of joint $1R$ and $1'R$ perceptions, it is necessary to slide from
statements about actual perceived results of actual experiments to
possible perceived results of experiments that were not actually
performed.  This is to extend the peculiar but undeniable ability of
consciousness to experience the particularity of a correlation in an
actual individual case to a hypothetical ability to experience the
fictional particularity of a correlation in a fictional case, and to
impose a consistency on the actually and fictionally perceived
particularities.  

It is hard to see how to make this compelling, unless what
consciousness is directly perceiving are actual correlata underlying
all the correlations.  If these had physical reality in an individual
case, locality would indeed require the value of a correlatum in one
subsystem to be the same, regardless of what local operations were
performed on the other subsystem.  But since quantum mechanics is
about correlations that exist without correlata, such an argument does
not work as a demonstration of nonlocality.\fn

\bigskip

There is another tradition of nonlocality arguments, starting with the
very first version of Bell's theorem which tests whether all the
correlations between currently non-interacting and far-apart
subsystems can be explained in terms of information commonly available
to the subsystems at the time of their last interaction.  This
``common-cause'' explanation for correlation assumes that it makes
sense to condition all joint subsystem distributions on the detailed
features of such hypothetical common information.  One then imposes
some reasonable locality conditions on these hypothetical conditional
distributions and shows that the resulting forms imply certain
inequalities that are inconsistent with the joint distributions given
by quantum mechanics.

From the perspective of the IIQM, if the pair of systems is completely
isolated from the rest of the world, such a conditioning on common
information is highly problematic, independent of the subsequent
imposing of locality conditions on such conditional distributions.
Refining the subsystem joint distributions according to ``conditions''
at the source, makes little sense from the perspective of the SSC
Theorem, which assures us that the correlations contain in themselves
complete information about the physical reality (encoded in the state)
of the two-subsystem system.  Such a refinement would grant physical
reality to further features of the correlations going beyond what is
contained in their joint (pure) state.  The only thing such arguments
show to be nonlocal is any such supplementation of the quantum
mechanical description.  Indeed, that was how Bell put it in his first
paper, and for some time thereafter the theorem was viewed not as a
proof that the physical world is nonlocal, but only as a nonlocality
proof for any hidden variables theory underlying the correlations.

It is, to be sure, a remarkable fact that the common-cause explanation
for correlation between non-interacting subsystems fails when applied
to quantum correlations, but this ought to be understood in terms of
the broader (equally remarkable) fact that correlation and only
correlation constitutes the full content of physical reality.  

\bigskip\leftline{{\bf XI. Comments on other
approaches.}}\nobreak\medskip\nobreak I first encountered the view
that correlations are fundamental and irreducible when I heard it
advocated as the proper way to think about Einstein--Podolsky--Rosen
(EPR) correlations, in talks by Paul Teller\fn and Arthur Fine.\fn It
did not then occur to me that this might be the proper way to think
about much more general correlations, but it should have, since this
is an important part of Bohr's reply\fn to EPR.\fn Nor did it occur to
me that objective reality might consist {\it only\/} of correlations
until I heard Lee Smolin\fn sketch an approach to quantum mechanics
that treated {\it symmetrically\/} a physical system and the world
external to that physical system.  Shortly thereafter I received a
beautiful paper from Carlo Rovelli\fn arguing from a very different
point of view that quantum states were expressions of relations
between subsystems.  Recently Gyula Bene\fn has written interestingly
along these lines.

This general attitude towards quantum states --- that the information
they contain is necessarily {\it relational\/} --- goes at least back
to Everett's original ``relative--state'' formulation of quantum
mechanics.\fn What is special to the IIQM
is (a) its insistence, justified by the SSC Theorem, on replacing
all talk about quantum {\it states\/} with talk about {\it subsystem
correlations\/}, (b) its insistence that {\it all\/} correlations
among subsystem observables for {\it all\/} resolutions into subsystems
have {\it joint\/} validity --- simultaneous physical reality, if you will,
and (c) its insistence that the {\it correlata\/} that underly those
correlations lie beyond the descriptive powers of physical science or,
equivalently, that although all subsystem {\it joint\/} distributions
are meaningful the corresponding {\it conditional\/} distributions are
not.

The IIQM evokes the Everett interpretation in stressing that a
measurement is nothing more than a particular kind of
interaction between two particular types of subsystems, designed to
yield a particular kind of correlation, and in stressing the fact that
a system $S_1$ that has non-trivial external correlations with a system
$S_2$, has no pure state of its own, even when the joint system $S =
S_1 + S_2$ is in a pure state $\k{\Psi}$.  The IIQM assigns a
fundamental status to the reduced density matrix of $S_1$ as the
complete embodiment of all its internal correlations.  Everett, on the
other hand, characterizes $S_1$ by a multitude of pure states, each
conditional on the assignment of an (almost arbitrary) pure state to
$S_2$.  Specifically, if $$P = \k{\chi}\b{\chi}\eq(PPP)$$ is a projection
operator on any pure state $\k{\chi}$ of $S_2$ and $$\b{\Psi}P\k{\Psi}
\neq 0,\eq(NNN)$$ then one easily establishes that there is a unique
pure state $\k{\phi}$ of $S_1$ for which the mean value of any
observable $A$ of $S_1$ is given by $$ \b{\phi}A\k{\phi} =
\b{\Psi}AP\k{\Psi}/\b{\Psi}P\k{\Psi}.\eq(relstate)$$ Everett calls
$\k{\phi}$ the state of $S_1$ {\it relative to\/} $\k{\chi}$ being the
state of $S_2$.

According to the IIQM Everett's relative states have no physical
significance, because the internal correlations of the
subsystem $S_1$ in the relative state $\k{\phi}$ are given by a
distribution that is {\it conditioned\/} on the other subsystem $S_2$
being in the state $\k{\chi}$.  While the {\it correlations\/} between
arbitrary observables of $S_1$ and the observable $P =
\k{\chi}\b{\chi}$ of $S_2$, or the corresponding {\it joint\/}
distributions, do have physical reality, the {\it conditional\/}
distribution for $S_1$ obtained by conditioning on $P$ {\it having the
value\/} 1 in $S_2$ does not.  As discussed in Section IX, one cannot
condition on the values of correlata, because such values have no
physical reality.  Thus Everett's relative states of a subsystem give
rise to internal correlations for that subsystem that are specified by
conditional distributions that have no physical meaning in the IIQM.
It is the insistence on the simultaneous reality of all these {\it
conditional\/} distributions that sends one off into the
cloud-cuckoo-land of many worlds.

Christopher Fuchs has suggested\fn that the distinction between the many worlds
interpretation and the ``correlations without correlata'' of the IIQM, is
most succinctly expressed by characterizing many worlds as {\it
correlata without correlations\/}.  In the many worlds interpretation
particular individual values of physical properties exist in
(over)abundance; but the problem of relating probabilities to the
branching of the worlds of different correlata has not been
satisfactorily resolved, in spite of many efforts going all the way
back to Everett's original paper.

There has also been a venerable tradition of talk about consciousness
and quantum physics, almost from the beginning.  My own talk is
closest to that which gives consciousness the power of ultimately
``reducing the wave packet.'' The difference is that the IIQM does not
speak of wave packet reduction at all, because if physical reality
consists only of correlations, nothing physically real ever changes
discontinuously.  To be sure a vestige of this point of view is
retained in my warnings to separate the problem of our mysterious
ability directly to perceive the particularity of our own correlation
with another macroscopic system from the problem of understanding
quantum mechanics.  But as noted in Section IV, the IIQM takes the
view that this ability poses a very hard problem about the nature of
our consciousness which ought not to be confused with the merely hard
problem of understanding the nature of quantum mechanics as applied to
a world devoid of consciousness.\fn

This point of view toward consciousness is in sharp contrast to a more
recent tradition, which tries to find an explanation for consciousness
based on quantum physics.\fn The IIQM takes quite the opposite
position, that consciousness experience goes beyond anything physics
is currently (and perhaps ever) capable of coming to grips with.

Two other interpretive schemes --- the modal interpretations\fn and
the consistent histories approach\fn --- also dethrone measurement.  Both can be distinguished
from the approach described here in terms of how they treat
correlations and correlata. Modal interpretations grant reality to
more than just relational quantities, at the price of restricting this
stronger reality to very special circumstances.  Subsystem
correlations {\it and\/} the associated correlata can be real provided
there are just two subsystems, and provided the correlations have the
strong form
\(joint).  This is made interesting by the Schmidt (polar)
decomposition theorem,\fn which guarantees that the state of any two-subsystem
system leads to such correlations for some choice of the two subsystem
observables.  But it leaves the status of other observables up in the
air, is embarrassed when the Schmidt decomposition of the two
subsystem state is not unique, and has nothing to say about three or
more subsystems.  

The consistent histories interpretation of quantum mechanics applies
to time-depend\-ent as well as equal-time correlations.  In contrast to
the IIQM, consistent historians are not at [Ball shy about dealing with
the correlata that underly a given set of correlations.  They gain
this interpretive flexibility by insisting that any talk about either
correlations or correlata must be restricted to sets of observables
singled out by certain quite stringent consistency conditions.  Thus
in the example of Section VII consistent historians may speak of
the correlations {\it and\/} the correlata for the observables 1 and
$1'$ or those for 1 and $2'$ or those for 2 and $1'$ or those for 2
and $2'$.  But they are forbidden to combine features of all these cases
into a single description.  These various incompatible descriptions
constitute mutually exclusive ``frameworks'' for describing a single
physical system.

I view the consistent histories interpretation as a formalization and
extension of Bohr's doctrine of complementarity.\fn The consistent
historians liberate complementarity from the context of
mutually exclusive experimental arrangements, by stating the
restrictions in terms of the quantum mechanical formalism itself,
without any reference to measurement.  This enables one within a given
framework to contemplate what {\it is\/} whether or not anything has
actually been {\it measured\/} --- indeed measurements in the
consistent histories interpretation (as in the IIQM and the Everett
interpretation) are simply a special case in which some of the
subsystems function as apparatuses.

The price one pays for this liberation is that the paradoxical quality
of complementarity is stripped of the protective covering furnished by
Bohr's talk of mutually exclusive experimental arrangements, and laid
bare as a vision of a single reality about which one can reason in a
variety of mutually exclusive ways, provided one takes care not to mix
them up.  Reality is, as it were, replaced by a set of complementary
representations, each including a subset of the correlations and their
accompanying correlata.  In the consistent histories interpretation it
is rather as if the representations have physical reality but the
representata do not.

The IIQM, in contrast, allows one to contemplate together {\it all\/}
subsystem correlations, associated with {\it all\/} complementary sets
of subsystem observables.  In justification of treating all such
correlations as simultaneously real one notes that quantum mechanics
allows one, given the state of the global system, to calculate
together the values of all such correlations; that the joint (but not
the conditional) distributions arrived at in this way are all mutually
consistent; and that quantum mechanics insures that the catalog of all
such joint subsystem distributions completely pins down the global
state.  The IIQM achieves this capability by denying to physics the
possibility of dealing with the individual correlata at all.

Whether this is a fatal defect of the IIQM, whether it is a
manifestation of the primitive state of our thinking about objective
probability, or whether it is a consequence of the inability of
physics to encompass conscious awareness, remains to be explored.

\bigskip

\leftline{{\bf XII. A few final remarks}}\nobreak\medskip\nobreak At
the risk of losing the interest of those who (like myself) read only
the first and last Sections before deciding whether the rest is worth
perusing, I conclude with some brief comments about loose ends.

As noted at the beginning, what I have been describing is more an
attitude toward quantum mechanics than a systematic interpretation.
The only proper subject of physics is how some parts of the world
relate to other parts.  Correlations constitute its entire content.
The actual specific values of the correlated quantities in the actual
specific world we know, are beyond the powers of physics to
articulate.  The answer to the question ``What has physical reality?''
depends on the nature of ``what''.  The answer is ``Everything!'' if
one is asking about correlations among subsystems, but ``Nothing!'' if
one is asking about particular values for the subsystem correlata.  

This alters the terms of the traditional debates.  Traditionally
people have been asking what {\it correlata\/} have physical reality.
The many different schools of thought differ by answering with many
different versions of ``Some'' while the IIQM answers ``None!'' The
question of what {\it correlations\/} have physical reality, which the IIQM
answers with ``All!'' has not, to my knowledge, been asked in this
context.  While I maintain that abandoning the ability of physics
to speak of correlata is a small price to pay for the recognition that
it can speak simultaneously and consistently of all possible
correlations, there remains the question of how to tie this wonderful
structure of relationships down to anything particular, if physics
admits of nothing particular.  

At this stage I am not prepared to offer an answer, beyond noting that
this formulates the conceptual problem posed by quantum mechanics in a
somewhat different way, and suggesting that there may be something to
be learned by thinking about it along these lines.  I suspect our
unfathomable conscious perceptions will have to enter the picture, as
a way of updating the correlations.  To acknowledge this is not to
acknowledge that ``consciousness collapses the wave-packet''.  But it
is to admit that quantum mechanics does not describe a world of
eternally developing correlation (described by ``the wave-function of
the universe''), but a phenomenology for investigating what kinds of
correlations can coexist with each other, and for updating current
correlations and extrapolating them into the future.  This
phenomenology applies to any system that can be well approximated as
completely isolated.

A skeptic might object that the problem of how to update correlations
is nothing more than the measurement problem, under a new name.
Perhaps it is, but at least the problem is posed in a new context: how
are we to understand the interplay between correlation as the only
objective feature of physical reality and the absolute particularity
of conscious reality?  Is something missing from a description of
nature whose purpose is not to disclose the real essence of the
phenomena but only to track down relations between the manifold
aspects of our experience?  Is this a shortcoming of our description
of nature or is it a deep problem about the nature of our experience?

Besides ``measurement'' John Bell\fn also disapproved of the word ``system'' --- a word I have
used uncritically more than a hundred times (not counting
``subsystem'', which occurs even more often).  If the purpose of
physics is to track down relations between the manifold aspects of our
experience, then there is nothing wrong in leaving the specifications
of the systems to us ourselves, however we manage to do it ---
sometimes by direct conscious perception, sometimes by deductions from
what we have learned from the correlations we have managed to induce
between the systems we can perceive and the ones we cannot.  Admitting
``system'' to the proper vocabulary of physics is not the same as
admitting ``correlata'' --- the (physically inacessible) particular
values of the quantifiable properties of an individual system.  

By acknowledging that in our description of nature the purpose is not
to disclose the real essence of the phenomena, we free ourselves to
construct from the manifold aspects of our experience formal
representations of the systems we want to talk about.  We have learned
how to express their possible correlations by an appropriate state
space, and the evolution of those correlations by an appropriate
Hamiltonian.  By setting aside ``the real essence of the phenomena''
we also acquire the ability to replace the befuddling spectre of an
endlessly branching state of the universe --- as disturbing in the
self-styled down-to-earth Bohmian interpretation as it is in the
wildest extravagances of the many worlds interpretation --- with a
quantum mechanics that simply tells us how we can expect some of the
manifold aspects of our experience to be correlated with others.
While this may sound anthropocentric, it is my expectation that {\it
anthropos\/} can be kept out of everything but the initial and final
conditions, and often (but not always) even out of those.

But this remains to be explored.

\bigskip \leftline{{\bf Acknowledgments}} \medskip If this view of
quantum mechanics has acquired more coherence since its first
appearance, much of this is due to responses my ``Ithaca
Interpretation'' essay elicited, and to reactions to innumerable
earlier drafts of the present essay.  For such thoughtful criticisms I
am indebted to Leslie Ballentine, Gilles Brassard, Rob Clifton,
Michael Fisher, Christopher Fuchs, Sheldon Goldstein, Kurt Gottfried,
David Griffiths, Robert Griffiths, Yuri Orlov, Abner Shimony, William Wootters,
and several anonymous reviewers of a proposal to the National Science
Foundation, which I am nevertheless pleased to be able to thank for
supporting this work under Grant No.~PHY9722065.  

\bigskip \leftline{{\bf Appendix A. The SSC Theorem: Subsystem
correlations determine the state}}\nobreak\medskip\nobreak Given a
system $\s = \s_1 + \s_2$ with density matrix $W$, then $W$ is
completely determined by the values of tr$W\,A\x B$ for an appropriate
set of observable pairs $A$, $B$, where $A = A\x 1$ is an observable
of subsystem $\s_1$ and $B = 1\x B$ is an observable of subsystem
$\s_2$. The proof is straightforward:

Give the state spaces for $\s_1$ and $\s_2$ orthonormal bases of
states $\k{\psi_\mu}$ and $\k{\phi_\alpha}$, respectively. Let the
$A$'s consist of the hermitian operators on $\s_1$
$$A_r^{(\mu\nu)} = \fr1/2\bigl(\k{\psi_\mu}\b{\psi_\nu} +
\k{\psi_\nu}\b{\psi_\mu}\bigr)\eq(A1)$$ and
$$A_i^{(\mu\nu)} = \fr1/{2i}\bigl(\k{\psi_\mu}\b{\psi_\nu} -
 \k{\psi_\nu}\b{\psi_\mu}\bigr),\eq(A2)$$ and let the $B$'s consist of the
hermitian operators on $\s_2$
$$B_r^{(\alpha\beta)} = \fr1/2\bigl(\k{\phi_\alpha}\b{\phi_\beta} +
\k{\phi_\beta}\b{\phi_\alpha}\bigr)\eq(B1)$$ and
$$B_i^{(\alpha\beta)} = \fr1/{2i}\bigl(\k{\phi_\alpha}\b{\phi_\beta} -
 \k{\phi_\beta}\b{\phi_\alpha}\bigr).\eq(B2)$$ 

The states $\k{\psi_\mu,\phi_\alpha} = \k{\psi_\mu}\x\k{\phi_\alpha}$ are a
complete orthonormal set of states for the composite system $\s$, and
the density matrix $W$ for the entire system $\s$ is determined by its
matrix elements  $$\b{\psi_\nu,\phi_\beta} W
\k{\psi_\mu,\phi_\alpha} = \tr W\bigl(
\k{\psi_\mu,\phi_\alpha}\b{\psi_\nu,\phi_\beta}\bigr).\eq(matW)$$  
But this can be expressed entirely in terms of quantities of the form
$\tr W (A \x B)$ --- i.e. in terms of subsystem correlations: 
  $$\b{\psi_\nu,\phi_\beta} W
\k{\psi_\mu,\phi_\alpha} =$$
$$\tr W\bigl(\k{\psi_\mu,\phi_\alpha}\b{\psi_\nu,\phi_\beta}\bigr)
= \tr W \Bigl(\bigl(A_r^{(\mu\nu)} + iA_i^{(\mu\nu)}\bigr)\x
              \bigl(B_r^{(\alpha\beta)} + i  B_i^{(\alpha\beta)}\bigr)
\Bigr) = $$
$$\tr W \bigl(A_r^{(\mu\nu)}\x B_r^{(\alpha\beta)}\bigr)
- \tr W  \bigl(A_i^{(\mu\nu)}\x B_i^{(\alpha\beta)}\bigr) $$ $$ 
+ i\tr W  \bigl(A_r^{(\mu\nu)}\x B_i^{(\alpha\beta)}\bigr)
+ i\tr W  \bigl(A_i^{(\mu\nu)}\x B_r^{(\alpha\beta)}\bigr).\eq(done)$$
Thus the values of the subsystem correlations between all the $A$'s
and $B$'s are enough to determine all the matrix elements of $W$ in a
complete set of states for the total system $\s$, and hence they are
enough to determine the density matrix $W$ for the total system.  

This proof straightforwardly generalizes to a system $\s = \s_1\+\cdots\+\s_n$
composed of more than two subsystems: given any resolution of $\s$
into $n$ subsystems, the density matrix of $\s$ is entirely determined
by the correlations among appropriate observables belonging to those
subsystems.  In such cases the structure of quantum mechanics
guarantees the important fact that it doesn't matter whether we pin
down the density matrix, for example, of $\s = \s_1 \+ \s_2 \+ \s_3$
from correlations between observables of $\s_1$ with observables that
act globally on $\s_2\+\s_3$, or from correlations between observables
of $\s_3$ with observables acting globally on $\s_1\+\s_2$, or from
tripartite correlations among observables acting only on the three
subsystems.

Thus the density matrix of a composite system determines all the
correlations among the subsystems that make it up and, conversely,
{\it the correlations among all the subsystems completely determine
the density matrix for the composite system they make up.\/} The
mathematical structure of quantum mechanics imposes constraints, of
course, on what those correlations can be --- namely they are
restricted to those that can arise from some global density matrix.
The particular form of that density matrix is then completely pinned
down by the correlations themselves. 

That the correlations cannot be
more general than that is the content of Gleason's
Theorem.\fn It would be interesting to explore the extent to which the
underlying structure of probabilities assigned to subspaces of a
Hilbert space on which Gleason's Theorem rests is itself pinned down
by the requirement of consistency among the different possible resolutions
of a system into subsystems.

\bigskip
\leftline{{\bf Appendix B. The external correlations of a
system}}\nobreak\leftline{{\bf are necessarily trivial if and only if
its state is pure}}\nobreak\medskip\nobreak We first show that if the
state of a subsystem $\s_1$ is pure, i.e. if its density matrix $W_1$
is a one-dimensional projection operator, $$W_1 = P_\phi =
\k{\phi}\b{\phi},\eq(W1)$$ then the density matrix $W$ of any larger
system system $\s = \s_1 + \s_2$ containing $\s_1$ as a subsystem must
be of the form $$W = P_\phi \x W_2, \eq(W)$$ and therefore all
external correlations of $\s_1$ are trivial.

This is easily established in the representation in which $W$ is
diagonal:  $$ W =
\sum_i w_i \k{\Psi^i}\b{\Psi^i},\eq(diag)$$ where the weights $w_i$
are non-negative.  If the reduced density matrix $W_1$ for $\s_1$ has
the form \(W1), then its diagonal elements $\b{\phi'}W_1\k{\phi'}$
must vanish for any state $\k{\phi'}$ in the state space of $\s_1$
orthogonal to $\k{\phi}$; i.e. if the $\k{\chi_n}$ are any orthonormal
basis for the state space of $\s_2$, then  $$0 =
\sum_n\b{\phi',\chi_n}W\k{\phi',\chi_n} =
\sum_{i,n}w_i|\ip{\phi',\chi_n}{\Psi^i}|^2.\eq(vanish)$$ Since the
$w_i$ are non-negative, for every non-zero $w_i$ we have $$0 =
\ip{\phi',\chi_n}{\Psi^i}.\eq(vanish1)$$ Now if the
$\k{\phi_j}$ are an orthonormal basis for the state space of $\s_1$
with $\k{\phi_1} = \k{\phi}$, then each $\k{\Psi^i}$ appearing in the
expansion \(diag) of $W$ is of the form $$\k{\Psi^i} =
\sum_{j,n}\k{\phi_j,\chi_n}\ip{\phi_j,\chi_n}{\Psi^i}\eq(fino)$$ It
follows from \(vanish1) that $$\k{\Psi^i} =
\sum_n\k{\phi,\chi_n}\ip{\phi,\chi_n}{\Psi^i} =
\k{\phi}\x\sum_n\k{\chi_n}\ip{\phi,\chi_n}{\Psi^i}.\eq(fini)$$ This
(with the form \(diag) of $W$) shows that $W$ is indeed of the form
\(W).

Conversely, if the state of $\s_1$ is mixed, then its density matrix
has the form $$W_1 = \sum p_i\k{\phi_i}\b{\phi_i}\eq(mixed)$$ where the
states $\k{\phi_i}$ are orthonormal and at least two of the $p_i$
(which we can take to be $p_1$ and $p_2$) are non-zero.  This density
matrix can arise if $\s_1$ is a subsystem of a larger system $\s =
\s_1 + \s_2$ with pure-state density matrix $$W = \k\Psi\b\Psi
\eq(WW),$$ where the state $\k\Psi$ is given by $$\k\Psi = \sum_i
\sqrt{p_i}\k{\phi_i}\x\k{\chi_i},\eq(Psi)$$ and the $\k{\chi_i}$ are
an orthonormal set of states for $\s_2$.  If observables $A_1$ and
$A_2$ are defined for each subsystem by $$A_1 = \k{\phi_1}\b{\phi_2} +
\k{\phi_2}\b{\phi_1} \eq(A1)$$ and $$A_2 = \k{\chi_1}\b{\chi_2} +
\k{\chi_2}\b{\chi_1} \eq(A2)$$ then $A_1$ and $A_2$ are non-trivially
correlated, since
$$\tr W A_1\x A_2 = 2\sqrt{p_1p_2}\eq(nontrivo)$$~
but $$\tr W A_1\x 1 = 0; \ \tr W\ 
1\x A_2 = 0.\eq(nontriv)$$

\bigskip
\leftline{{\bf Appendix C. Glossary of
terms}}\nobreak\medskip\nobreak{\parindent=0pt \parskip = 6
pt\baselineskip = 14pt
\nobreak{\sl Internal correlations.} The internal correlations of a
system are the correlations prevailing among any of its subsystems.

{\sl External correlations.} The external correlations of a system are
those it has with other systems which together constitute the
subsystems of a larger system.

{\sl Trivial correlations.} Subsystem correlations arising from joint
probabilities that are products of subsystem probabilities.

{\sl Non-trivial correlations.} Correlations that are not trivial ---
i.e. in which the mean of some products differs from the product of the
means.

{\sl State.}  The state of a system is the complete set of all its
internal correlations.  These are concisely encoded in its density
matrix.

{\sl SSC Theorem.}  The theorem on the sufficiency of subsystem
correlations for a complete determination of the quantum state of a
composite system.  It is stated and proved in Appendix A; see also
Ref.~17.

{\sl Pure state.}  The state of a system whose density matrix is a
one-dimensional projection operator.  Or, equivalently, the state of a
system that has no nontrivial external correlations.

{\sl Mixed state.}  The state of a system whose density matrix is not
a one-dimensional projection operator.  Or, equivalently, the state of
a system that can have nontrivial external correlations.

{\sl Dynamically isolated system.} A system that has no external
interactions.

{\sl Completely isolated system.} A system that has no external
interactions or correlations.

{\sl Specimen.} A subsystem (usually microscopic) that we wish to
learn something about. 

{\sl Apparatus.} A macroscopic subsystem we dynamically correlate
with a specimen. 

{\sl Measurement.} The dynamical process by which the correlations
between a specimen and an apparatus are brought into the particular
canonical form \(joint).

{\sl Physical locality.} The fact that the internal correlations of a
dynamically isolated system do not depend on any interactions
experienced by other systems external to it. 

{\sl Metaphysical locality.} The requirement (often violated) that the
external correlations of a dynamically isolated system should make
sense in terms of internal correlata.   

{\sl Correlatum.} The particular value of a property of an individual
system (represented in the formalism by a particular eigenvalue of the
corresponding hermitian operator). According to the IIQM correlations
among the correlata of different subsystems have physical reality but
the correlata themselves do not.

{\sl Physical reality.} That whereof physics can speak.  For example the
physical reality of {\it blue\/} includes a certain class of Fourier
decompositions of the radiation field, and the excitations in the
retina produced by fields with such Fourier decompositions, and the
signals transmitted by such excitations to the visual cortex.  

{\sl Reality.} Physical reality plus that on which physics is silent,
its conscious perception.  For example for me the reality of {\it blue}
consists of its physical reality augmented by my accompanying sensation
of {\it blueness\/}.

{\sl IIQM.} ``The Ithaca Interpretation of Quantum Mechanics'' --- the
constellation of ideas put forth above, more accurately characterized
as ``{\it An\/} Ithaca Interpretation of Quantum Mechanics''.

}

\vfil\eject 

\baselineskip=15pt
\leftline{{\bf Notes and References.}}
\parindent= 0pt
\parskip = 8pt

1.  Notes for a lecture given at the Symposium in Honor of Edward
M.~Purcell, Harvard University, October 18, 1997.  Published in
Am. J. Phys., {\bf 66}, 753-767 (1998).

2.  Introductory Survey to {\it Atomic Theory and
the Description of Nature\/}, Cambridge, 1934, p. 18.  Reprinted in
{\it Niels Bohr, Collected Works\/}, vol. 6, North Holland (1985),
p.296.

3.  N.~David Mermin, ``The Ithaca Interpretation of Quantum
Mechanics,'' to appear in Pramana; a version can be found in quant-ph
9609013, Los Alamos e-Print archive at xxx.lanl.gov (1996).  The
nomenclature was intended to indicate a resemblance to the body of
interpretational lore named after a grander city in northern Europe,
and also, by its geographic modesty and lack of descriptive content,
to suggest that what was being promulgated was not so much a logical
foundation for the subject as a philosophical perspective or
pedagogical approach.  It does not imply that others in Ithaca share
these views, or that others outside of Ithaca have not expressed
similar thoughts.  None of the ingredients of the IIQM are novel, but
I have cooked them together into a somewhat different stew.

4. These are examples of the kinds of terms or distinctions that have
a special character in the IIQM, and are collected together in
Appendix C.

5.  D.~Bohm and D.~J.~Hiley, {\it The Undivided
Universe}, Routledge, New York, 1993.

6. For a recent review and
a set of references, see for example Philip Pearle, ``Wavefunction
Collapse Models With Nonwhite Noise'', in {\it Perspectives on Quantum
Reality\/}, Rob Clifton, ed., Kluwer, Dordrecht, 1996, pp.~93-109.

7.A technical point: In
taking the state space of the system to be a product of subsystem
state spaces I am restricting the discussion to cases where the
significant manifestations of quantum mechanical indistinguishability
of particles --- the symmetry or anti-symmetry of many particle wave
functions --- are limited to the constituents of the individual
subsystems.  This is conventional (though rarely noted) in discussions
of the foundations of quantum mechanics.  Thus in discussing the
measurement of the spin of an atom, one does not antisymmetrize the
combined wave function for the atom-apparatus system over, for
example, the electronic variables occurring in both subsystems. The
overlap between atomic and apparatus electronic wave functions is
taken to be zero.  Since the major conceptual problems posed by
quantum mechanics --- nonlocality and the measurement problem --- are
present even in the quantum mechanics of distinguishable particles,
this simplification does not appear to evade essential features.  A
more rigorous approach will probably require a field-theoretic
formulation.  There seems no point in trying to cross that bridge
unless one can first cross the simpler one attempted here.  In a
similar vein, I also consider here only non-relativistic quantum
mechanics, because the conceptual problems are already present in the
non-relativistic theory.  A treatment of relativistic quantum
mechanics will also require a field theoretic reformulation.

8. This is spelled out more explicitly at the beginning of
Section IX.

9. For example the
joint distribution for electron position and proton position in a
hydrogen atom exists simultaneously with the joint distribution for
electron momentum and proton position, even though the position and
momentum of the electron do not have joint physical reality or a
meaningful joint distribution of their own.  And both the
position-position and momentum-position distributions return the same
distribution for the proton position, when the electronic variables
are integrated out.

10. The essential role of objective probability in the quantum
mechanical description of an individual system was stressed by Popper,
who used the term ``propensity''. See Karl Popper, {\it Quantum Theory
and the Schism in Physics,\/} Rowman and Littlefield, Totowa, New
Jersey, 1982.  Heisenberg may have had something similar in mind with
his term ``potentia''.  While I agree with Pop\-per that quantum
mechanics requires us to adopt a view of probability as a fundamental
feature of an individual system, I do not believe that he gives
anything like an adequate account of how this clears up what he called
the ``quantum mysteries and horrors''.  See N.~David Mermin, ``The
Great Quantum Muddle,'' Philosophy of Science {\bf 50}, 651-656
(1983); reprinted in {\it Boojums All the Way Through\/}, Cambridge
(1990), pps.190-197.

11. Wolfgang Pauli,
``Probability and physics'', in {\it Writings on Physics and
Philosophy\/}, Springer-Verlag, New York (1994), 43-48.

12. I comment further on the Everett
interpretation, which was subsequently transformed into the many-worlds
interpretation, in Section XI.

13. Einstein was apparently resigned to the inaccessibility of \now\
to physics.  According to Carnap (``Intellectual
Autobiography,'' in P.~A.~Schilpp (ed.), {\it The Philosophy of Rudolf
Carnap\/}, Open Court, LaSalle Illinois (1963), pp. 37-38) in a
conversation in the early 1950's ``Einstein said that the problem of
the Now worried him seriously.  He explained that the experience of
the Now means something special for man, something essentially
different from the past and the future, but that this important
difference does not and cannot occur within physics.  That this
experience cannot be grasped by science seemed to him a matter of
painful but inevitable resignation.''  This is particularly
interesting in view of Einsteins notorious unwillingness to extend his
resignation over the inability of physics to deal with the special
character of \now, to its inability to deal with the special character
of correlata underlying the quantum correlations.

14.  To my surprise, this point --- a banality among philosophers, who
speak of {\it qualia\/} --- is extremely hard, if not impossible, to
put across to some physicists.  I have sometimes managed to do it by
citing a theory I had as a child to account for the fact that
different people have different favorite colors.  My idea --- a kind
of chromo-aesthetic absolutism --- was that there was, in fact, only
one most pleasurable color sensation, but the reason {\it your\/}
favorite color was blue while {\it mine\/} was red was that the
sensation you experienced looking at blue objects was identical to the
sensation I experienced looking at red ones. I recently found
precisely this example (complete to the choice of colors --- only
``you'' and ``me'' are interchanged) on a list of possibly meaningless
questions in P.~W.~Bridgman, {\it The Logic of Modern Physics\/},
Macmillan (1927), p.~30.

15.  For an example see the discussion of quantum nonlocality in
Section X.

16. For an engaging discussion of these issues and many
references, see Euan Squires, {\it Conscious Mind in the Physical
World\/}, Adam Hilger, Bristol and New York, 1990.

17.  S. Bergia, F.  Cannata, A. Cornia, and R. Livi, "On the actual
measurability of the density matrix of a decaying system by means of
measurements on the decay products", Foundations of Physics 10,
723-730 (1980).  See also W.~K.~Wootters, in {\it Complexity, Entropy and the
Physics of Information}, W.~H. Zurek, ed., Addison-Wesley, Redwood
City, California, 1990, pp.~39-46.

18. J.~S.~Bell, ``Against measurement,''Physics World, August, 1990,
33-40.  This critique elicited interesting rejoinders from Rudolf
Peierls (``In defense of measurement'', Physics World, January, 1991,
19-20) and Kurt Gottfried (``Does quantum mechanics carry the seeds of
its own destruction?'', October, 1991, 31-40.) Bell's death deprived
us of his response.

19.  I defer to Section VIII any discussion of ``the measurement
problem'' --- the constellation of issues arising in the context of
``wave-packet collapse''.

20.  FAPP = For all practical purposes. See J. S. Bell, {\it
loc.~cit.}

21.  The exact absence of interference effects for any observables
associated entirely with either the system or the apparatus (i.e. of
the form $S\x 1$ or $1\x A$ for arbitrary system and apparatus
observables $S$ and $A$), is, of course, also directly evident from
the form \(F) of the post-measurement (pure) state of the total
specimen-apparatus system, in which the phases of the $\alpha_i$ still
appear.

22.  Hugh Everett, III, ``Relative-State Formulation of Quantum
Mechanics,'' Revs. Mod.  Phys.  {\bf 29}, 454-462 (1957).  Everett
says virtually nothing about many worlds except, perhaps, in a note
added in proof.  I discuss the relation of the IIQM to Everett's
relative-state formulation in Section XI.

23.  See N. David Mermin, ``Hidden
Variables and the Two Theorems of John Bell,'' Revs.~Mod.~Phys.~{\bf
65}, 803-815 (1993), especially Sec.~VII.

24.  Lucien Hardy, ``Quantum mechanics, local realistic
theories, and Lorentz-invariant realistic theories,'',
Phys.~Rev.~Lett.~{\bf 68}, 2981-2984 (1992).  The version of Hardy's
argument given here uses the notation in N. David Mermin,
``Quantum mysteries refined'', Am.~J.~Phys.~{\bf 62}, 880-897 (1994).

25. Nor is the absence of subsystem correlata a peculiarity of a small
class of specially contrived states.  Hardy has shown (in the
context of a ``nonlocality'' argument, but the theorems apply equally
well in the present context) that this state of affairs is generic,
holding for appropriate subsystem observables whenever a system $\s =
\s_1 + \s_2$ has {\it any\/} non trivial correlations between its
subsystems $\s_1$ and $\s_2$ (unless the individual subsystem
probabilities are completely random --- i.e.  unless the individual
subsystem density matrices are proportional to the unit matrix.)
See Lucien Hardy, ``Nonlocality for two particles without inequalities for
almost all entangled states,'' Phys.~Rev.~Lett.~{\bf 71}, 1665-1668
(1993).

26. John S.~Bell, ``On the
Einstein-Podolsky-Rosen paradox,'' Physics {\bf 1}, 195-200 (1964).

27. Most recently in Henry P.~Stapp, ``Nonlocal character of quantum
theory,'' Am. J. Phys. {\bf 65}, 300-304 (1997).

28.  I have given a detailed analysis of how Stapp's carefully constructed
derivation of nonlocality from the Hardy state can be used
to illuminate Bohr's reply to Einstein, Podolsky, and Rosen.  See
N.~David Mermin, ``Nonlocal character of quantum theory?'', submitted
to the American Journal of Physics. Stapp responds in ``Quantum
nonlocality'', simultaneously submited to the American Journal of
Physics.

29. Paul Teller, in {\it Philosophical
Consequences of Quantum Theory\/}, James T. Cushing and Ernan
McMullin, eds., Notre Dame Press, Notre Dame, Indiana, 1989, pp.
208-223.

30.  Arthur Fine, in {\it Philosophical
Consequences of Quantum Theory\/}, James T. Cushing and Ernan
McMullin, eds., Notre Dame Press, Notre Dame, Indiana, 1989,
pp.~175-194.

31.  Niels
Bohr, ``Can Quantum-Mechanical Description of Physical Reality Be
Considered Complete?'', Phys.~Rev.~{\bf 48}, 696-702 (1935).  One of
Bohr's points is that there is nothing new or unusual about EPR
correlations:  precisely the same kinds of correlations are set up in
the measurement process, and therefore there is no cause for alarm
because he has already straightened out that problem.

32.  Albert Einstein, Boris Podolsky, and Nathan Rosen, ``Can
Quantum-Mechanical Description of Physical Reality Be Considered
Complete?'', Phys.~Rev.~{\bf 47}, 777-780 (1935).

33.  This point of view is expressed in Lee Smolin, {\it The Life of
the Cosmos\/}, Oxford University Press (1997).

34. Carlo Rovelli, ``Relational
Quantum Mechanics'', International Journal of Theoretical Physics,
{\bf 35}, 1637-78 (1996)   See also quant-ph 9609002, Los
Alamos e-Print archive at xxx.lanl.gov (1996).

35. Gyula Bene, ``Quantum Reference systems: a new framework for
quantum mechanics, quant-ph/9703021, Los Alamos e-Print archive at
xxx.lanl.gov (1996), and to appear in Physica A; ``Quantum phenomena
do not violate the principle of locality -- a new interpretation with
physical consequences,'' quant-ph/9706043, Los Alamos e-Print archive
at xxx.lanl.gov (1996), submitted to Am.~J.~Phys.

36.  H.~Everett, {\it loc.~cit.\/}.  It was later swept
off into the many--worlds interpretation.

37. Christopher Fuchs, private communication.

38.  A similar attitude has been
expressed by Rudolf Peierls, {\it Surprises in Theoretical
Physics\/}, Princeton University Press, Princeton, New Jersey (1979),
p.~33: ``We are confident today that, if we could solve the
Schr\"odinger equation for all the electrons in a large molecule, it
would give us all the knowledge that chemists are able to discover
about it.$\,\ldots$Many people take it for granted that the same must be
true of the science of life.  The difficulty about how to formulate
the acquisition of information, which we have met, is a strong reason
for doubting this assumption.'' Even closer is the view of Robert
Geroch, ``The Everett interpretation'', Nous {\bf
18}, 617-633 (1984), p.~629: ``[W]hat must be accounted for$\ldots$
is, not the specific classical outcomes deemed to have occurred for a
specific experiment, but rather the general human impression that
classical outcomes do occur.  This problem may well be soluble, but is
probably beyond our present abilities; and, in any case, is basically
not a problem in quantum mechanics.''

39.  See, for example, Roger Penrose, {\it The Emperor's New Mind\/},
Oxford, New York, 1989, and {\it Shadows of the Mind\/}, Oxford, New
York, 1994; Henry Stapp, {\it Mind, Matter, and Quantum Mechanics},
Springer Verlag, New York, 1993.

40.  See, for example, Jeffrey Bub, {\it Interpreting the Quantum
World\/}, Cambridge, 1997 and Bas C.~Van Fraassen, {\it Quantum
Mechanics: An Empiricist View\/}, Clarendon Press, Oxford, 1991, and
references cited therein.

41.  For a recent formulation and references see Robert B.~Griffiths,
``Consistent Histories and Quantum Reasoning'', Phys.~Rev.~A{\bf 54},
2759-2774 (1996).

42.  See, for example, Asher Peres, {\it Quantum
Theory: Concepts and Methods\/}, Kluwer Academic, Dodrecht, 1993,
pp.123-126.

43.  Indeed, I believe its character would be clarified if consistent
historians were to characterize mutually exclusive families of
correlations and correlata not as {\it inconsistent\/} or {\it
incompatible\/} but as {\it complementary\/}. This terminology is
suggested on p.~162 of Roland Omn\`es, {\it The Interpretation of
Quantum Mechanics\/}, Princeton (1994).

44.  J.~S.~Bell, Physics World, loc.~cit.

45.  A.~M.~Gleason, ``Measures on the Closed Subspaces of a
Hilbert Space,'' Journal of Mathematics and Mechanics {\bf 6}, 885-893
(1957).

\bye